\documentclass[10pt,draft]{dis03}
\usepackage{epsf,amsmath}

\begin{document}
\title{SOLUTION OF THE BFKL EQUATION \\ AT NEXT-TO-LEADING ORDER
\thanks{This work has been supported by the UK Particle Physics and Astronomy Research Council (PPARC) (Posdoctoral Fellowship: PPA/P/S/1999/00446).}}

\author{AGUST{\'\i}N SABIO VERA\\\\
Cavendish Laboratory,
  University of Cambridge, Madingley Road,\\ CB3 0HE, Cambridge, U.K.}

\maketitle

\begin{abstract}
\noindent 
We solve the Balitsky--Fadin--Kuraev--Lipatov (BFKL) equation in
  the next--to--leading logarithmic approximation for forward scattering 
with all conformal spins using an iterative method . 
\end{abstract}

\section{Introduction}

The BFKL \cite{FKL} formalism resums large logarithms appearing in the Regge limit, where the center of mass energy $\sqrt{s}$ is large and
the momentum transfer $\sqrt{-t}$ fixed. The cross-section for the process $A+B \rightarrow A'+B'$ reads 
\begin{eqnarray}
\label{cross--section1}
\sigma(s) &=&\int 
\frac{d^2 {\bf k}_a}{2 \pi{\bf k}_a^2}
\int \frac{d^2 {\bf k}_b}{2 \pi {\bf k}_b^2} ~\Phi_A({\bf k}_a) ~\Phi_B({\bf k}_b)
~f \left({\bf k}_a,{\bf k}_b, \Delta = \ln{\frac{s}{s_0}}\right),
\end{eqnarray}
where $\Phi_{A,B}$ are the impact factors and $f\left({\bf
    k}_a,{\bf k}_b,\Delta\right)$ is the gluon Green's
function describing the interaction between two Reggeised gluons exchanged in
the $t$--channel with transverse momenta ${\bf k}_{a,b}$. We use the Regge scale $s_0 = \left|{\bf k}_a\right|\left|{\bf k}_b\right|$. In the
leading logarithmic approximation (LLA) terms of the form
$\left(\alpha_s \Delta \right)^n$ are resummed. In the next--to--leading
logarithmic approximation (NLLA) \cite{Fadin:1998py} contributions of the type $\alpha_s \left(\alpha_s\Delta \right)^n$ are also taken into account.

The Green's function is the solution of an integral equation where radiative corrections
enter through its kernel. In this contribution we solve
this equation using an iterative method considering
the full kernel with scale invariant and running coupling terms. In this
approach we keep all the angular information from the BFKL evolution, solving
the equation for a general conformal spin without relying on any asymptotic
expansion. For more details see Ref. \cite{Andersen:2003an}.

\section{The BFKL equation in the NLLA}
\label{BFKL@NLLA}

The BFKL equation is written in terms of a Mellin transform in $\Delta$ space, i.e. 
\begin{eqnarray}
\label{Mellin}
f \left({\bf k}_a,{\bf k}_b, \Delta\right) 
&=& \frac{1}{2 \pi i}
\int_{a-i \infty}^{a+i \infty} d\omega ~ e^{\omega \Delta} f_{\omega} 
\left({\bf k}_a ,{\bf k}_b\right).
\end{eqnarray}
The BFKL equation in the NLLA in dimensional regularisation then reads 
\begin{eqnarray}
\omega f_\omega \left({\bf k}_a,{\bf k}_b\right) &=& \delta^{(2+2\epsilon)} 
\left({\bf k}_a-{\bf k}_b\right) + \int d^{2+2\epsilon}{\bf k}' ~
\mathcal{K}\left({\bf k}_a,{\bf k}'\right)f_\omega \left({\bf k}',{\bf k}_b 
\right),
\end{eqnarray}
with the kernel $\mathcal{K}\left({\bf k}_a,{\bf k}\right) = 2 \,\omega^{(\epsilon)}\left({\bf k}_a^2\right) \,\delta^{(2+2\epsilon)}\left({\bf k}_a-{\bf k}\right) + \mathcal{K}_r\left({\bf k}_a,{\bf k}\right)$ depending on the gluon Regge trajectory and a real emission component \cite{Fadin:1998py}.

We split the kernel $\mathcal{K}_r$ into a 
$\epsilon$--dependent  ($\mathcal{K}_r^{(\epsilon)}$) and a
$\epsilon$--independent ($\widetilde{\mathcal{K}}_r$) parts. To
 show the cancellation of the $\epsilon$ poles  
we split the integral over transverse phase space for
$\mathcal{K}_r^{(\epsilon)}$ into two regions separated by a small cut--off 
$\lambda$. We then approximate $f_\omega \left({\bf k}_a+{\bf k},{\bf k}_b\right) \simeq
f_\omega \left({\bf k}_a,{\bf k}_b\right)$ for $\left|{\bf k}\right| <
\lambda$. This $\lambda$--dependence is negligible for large $\left|{\bf k}_a\right|$. In this way we express the BFKL equation as 
\begin{eqnarray}
\omega f_\omega \left({\bf k}_a,{\bf k}_b\right) &=& \delta^{(2+2\epsilon)}
\left({\bf k}_a-{\bf k}_b\right) \\
&&\hspace{-3cm}+ \left\{2 \, 
\omega^{(\epsilon)} \left({\bf k}_a^2\right) + \int d^{2+2\epsilon}{\bf k} \, \mathcal{K}_r^{(\epsilon)} \left({\bf k}_a,{\bf k}_a+{\bf k}\right) \theta\left(\lambda^2-{\bf k}^2\right) \right\}f_\omega \left({\bf k}_a,{\bf k}_b\right)\nonumber\\
&&\hspace{-3cm}+ \int d^{2+2\epsilon}{\bf k} \, \left\{ \mathcal{K}_r^{(\epsilon)} \left({\bf k}_a,{\bf k}_a+{\bf k}\right) \theta\left({\bf k}^2-\lambda^2\right) + \widetilde{\mathcal{K}}_r \left({\bf k}_a,{\bf k}_a+{\bf k}\right) \right\} f_\omega \left({\bf k}_a+{\bf k},{\bf k}_b\right).\nonumber
\end{eqnarray}

In dimensional regularisation the gluon Regge trajectory \cite{Fadin:1998py} reads \footnote{$\beta_0 \equiv \frac{11}{3}N_c-\frac{2}{3}n_f$, $\bar{\alpha}_s \equiv \frac{\alpha_s (\mu) N_c}{\pi}$, $\alpha_s (\mu) = \frac{g_\mu^2}{4 \pi}$, $\mu$ is the $\overline{\rm MS}$ renormalisation scale.}
\begin{eqnarray}
2 \, \omega^{(\epsilon)} \left({\bf q}^2\right) &=& - \bar{\alpha}_s \frac{\Gamma (1-\epsilon)}{(4 \pi)^\epsilon} \left(\frac{1}{\epsilon}+\ln{\frac{q^2}{\mu^2}}\right) 
- \frac{\bar{\alpha}_s^2}{8}\frac{\Gamma^2(1-\epsilon)}{(4 \pi)^{2 \epsilon}}
\left\{\frac{\beta_0}{N_c}\left(\frac{1}{\epsilon^2}+\ln^2{\frac{q^2}{\mu^2}}\right)\right.\nonumber\\
&+& \left.\left(\frac{4}{3}-\frac{\pi^2}{3}+\frac{5}{3}\frac{\beta_0}{N_c}\right)\left(\frac{1}{\epsilon}+2\ln{\frac{q^2}{\mu^2}}\right)-\frac{32}{9}+2 \zeta(3)-\frac{28}{9}\frac{\beta_0}{N_c}\right\}.
\end{eqnarray}
The $\epsilon$-dependent part of the real emission kernel \cite{Fadin:1998py} is
\begin{eqnarray}
\mathcal{K}_r^{(\epsilon)} \left({\bf q},{\bf q}+{\bf k}\right) &=&
\frac{\bar{\alpha}_s \mu^{-2 \epsilon}}{\pi^{1+\epsilon}(4 \pi)^\epsilon}
\frac{1}{{\bf k}^2} \left\{1+\frac{\bar{\alpha}_s}{4}\frac{\Gamma(1-\epsilon)}{(4 \pi)^\epsilon}\left[\frac{\beta_0}{N_c}\frac{1}{\epsilon}\left(1-\left(\frac{{\bf k}^2}{\mu^2}\right)^\epsilon \right.\right.\right.\\
&&\hspace{-3.2cm}\left.\left.\left.\times \left(1-\epsilon^2 \frac{\pi^2}{6}\right) \right) 
+\left(\frac{{\bf k}^2}{\mu^2}\right)^\epsilon \left(\frac{4}{3}-\frac{\pi^2}{3}+\frac{5}{3}\frac{\beta_0}{N_c}+\epsilon\left(-\frac{32}{9}+14\zeta(3)-\frac{28}{9}\frac{\beta_0}{N_c}\right)\right)\right]\right\}.\nonumber
\end{eqnarray}
When we combine the trajectory with the integration of $\mathcal{K}_r^{(\epsilon)}$ over the phase space limited by the cut--off, the poles in $\epsilon$
cancel and we obtain
\begin{eqnarray}
\omega_0 \left({\bf q}^2,\lambda^2\right) &\equiv&\lim_{\epsilon \to 0} \left\{ 2\, \omega^{(\epsilon)}\left({\bf q}^2\right) + \int d^{2+2\epsilon}{\bf k} \,
\mathcal{K}_r^{(\epsilon)} \left({\bf q},{\bf q}+{\bf k}\right) 
\theta \left(\lambda^2-{\bf k}^2\right) \right\} \\
&&\hspace{-2.4cm}= - \bar{\alpha}_s \left\{\ln{\frac{{\bf q}^2}{\lambda^2}}
+ \frac{\bar{\alpha}_s}{4}\left[\frac{\beta_0}{2 N_c}\ln{\frac{{\bf q}^2}{\lambda^2}}\ln{\frac{\mu^4}{{\bf q}^2 \lambda^2}}+\left(\frac{4}{3}-\frac{\pi^2}{3}+\frac{5}{3}\frac{\beta_0}{N_c}\right)\ln{\frac{{\bf q}^2}{\lambda^2}}-6 \zeta(3)\right]\right\}.\nonumber
\end{eqnarray}
Using $\omega_0 \left({\bf q}^2,\lambda^2 \right) \equiv - \xi\left(\left|{\bf q}\right|\lambda\right) \ln{\frac{{\bf q}^2}{\lambda^2}} + \eta $, $\eta \equiv {\bar{\alpha}_s}^2 \frac{3}{2} \zeta (3)$ and 
$\xi \left({\rm X}\right) \equiv \bar{\alpha}_s +  
\frac{{\bar{\alpha}_s}^2}{4}\left[\frac{4}{3}-\frac{\pi^2}{3}+\frac{5}{3}\frac{\beta_0}{N_c}-\frac{\beta_0}{N_c}\ln{\frac{{\rm X}}{\mu^2}}\right]$ we can write the equation in the simple form
\begin{eqnarray}
\label{nll}
\left(\omega - \omega_0\left({\bf k}_a^2,\lambda^2\right)\right) f_\omega \left({\bf k}_a,{\bf k}_b\right) &=& \delta^{(2)} \left({\bf k}_a-{\bf k}_b\right)\\
&&\hspace{-4cm}+ \int d^2 {\bf k} \left(\frac{1}{\pi {\bf k}^2} \xi \left({\bf k}^2\right) \theta\left({\bf k}^2-\lambda^2\right)+\widetilde{\mathcal{K}}_r \left({\bf k}_a,{\bf k}_a+{\bf k}\right)\right)f_\omega \left({\bf k}_a+{\bf k},{\bf k}_b\right),\nonumber
\end{eqnarray}
where $\widetilde{\mathcal{K}}_r \left({\bf q},{\bf q}'\right)$
contains the full angular information of the BFKL evolution and can be found 
in Ref. \cite{Andersen:2003an}. 

\section{Iterative solution in the NLLA}
\label{Iterating}

We solve Eq. (8) using an iterative procedure in the $\omega$ plane 
similar to the one in \cite{LLAlimit} for the LLA. 
In the NLLA we take the renormalisation scale $\mu = \mu \left({\bf k}_a^2, {\bf k}_b^2\right)$ depending on the large scales in the scattering process. 
Our result reads \footnote{Using the notation $y_0 \equiv \Delta$.}
\begin{eqnarray}
\label{ours}
f({\bf k}_a ,{\bf k}_b, \Delta) 
&=& \exp{\left(\omega_0 \left({\bf k}_a^2,{\lambda^2},\mu\left({\bf k}_a^2,{\bf k}_b^2\right)\right) \Delta \right)}
\left\{\frac{}{}\delta^{(2)} ({\bf k}_a - {\bf k}_b) \right. \\
&&\hspace{-3cm}+ \sum_{n=1}^{\infty} \prod_{i=1}^{n} 
\int d^2 {\bf k}_i \left[\frac{\theta\left({\bf k}_i^2 - \lambda^2\right)}{\pi {\bf k}_i^2} \xi\left({\bf k}_i^2,\mu\left(\left({\bf k}_a+\sum_{l=0}^{i-1}{\bf k}_l\right)^2,{\bf k}_b^2\right)\right) \right. \nonumber\\
&&\hspace{1cm}+\left. \widetilde{\mathcal{K}}_r \left({\bf k}_a+\sum_{l=0}^{i-1}{\bf k}_l,
{\bf k}_a+\sum_{l=1}^{i}{\bf k}_l,\mu \left(\left({\bf k}_a+\sum_{l=0}^{i-1}{\bf k}_l\right)^2,{\bf k}_b^2\right)\right)\frac{}{}\right]\nonumber\\
&& \hspace{-3cm} \times  
\int_0^{y_{i-1}} d y_i ~ {\rm exp}\left[\left(
\omega_0\left(\left({\bf k}_a+\sum_{l=1}^i {\bf k}_l\right)^2,\lambda^2,
\mu \left(\left({\bf k}_a+\sum_{l=0}^{i}{\bf k}_l\right)^2,{\bf k}_b^2\right)\right)\right.\right.\nonumber\\
&&\left.\left.\left.\hspace{-3cm}
-\omega_0\left(\left({\bf k}_a+\sum_{l=1}^{i-1} {\bf k}_l\right)^2,
{\lambda^2},\mu\left(\left({\bf k}_a+\sum_{l=0}^{i-1}{\bf k}_l\right)^2,{\bf k}_b^2\right)\right)\right) y_i\right] \delta^{(2)} \left(\sum_{l=1}^{n}{\bf k}_l 
+ {\bf k}_a - {\bf k}_b \right)\right\}, \nonumber
\end{eqnarray}
where we have made use of an 
inverse Mellin transform to write the final solution in energy space \cite{Andersen:2003an}.

\section{Conclusions}
\label{conclusions}

We have presented a method to solve the BFKL equation for forward scattering in the next--to--leading logarithmic approximation using the kernel in dimensional regularisation and
introducing a cut--off in phase space. This
allows us to write the solution in a compact form, Eq.~(\ref{ours}), suitable
for numerical studies, which will be presented in a future work. We keep the full angular information in our solution by solving the equation for any conformal spin. This will allow the study of spin--dependent observables in the NLLA.

In recent years there have been many studies of the behaviour of the gluon Green's function in the NLLA \cite{NLLpapers}. Work is in progress to understand the BFKL resummed gluon Green's function using this novel approach, and to quantify the effect of those terms related to the
running of the coupling \cite{running} compared to the scale invariant ones. 

Our ultimate goal is the calculation of
cross--sections in the NLLA. In principle, using this procedure it will be 
possible to
disentangle the structure of the final state allowing the study of, e.g.,
multiplicities, extending the work of~\cite{phenomenology} to the NLLA. 

\section*{Acknowledgements} 
I would like to thank Jeppe R. Andersen, my collaborator in this work, and the participants of 
DIS 2003 for their interest in our results, in particular, Jochen Bartels, 
Dimitri Colferai, Victor Fadin, Anatoly Kotikov, Albrecht Kyrieleis, Peter 
Landshoff, Lev Lipatov, Anna Stasto, Juan Terr{\' o}n and Gian Paolo Vacca.


\begin{thebibliography}{99} 

\bibitem{FKL}  L.\thinspace N.~Lipatov, Sov.\ J.\ Nucl.\ Phys.\ {\bf 23}, 338 (1976); V.\thinspace S.~Fadin, E.\thinspace A.~Kuraev and L.\thinspace N.~Lipatov, Phys.\ Lett.\ B {\bf 60}, 50 (1975), Sov.\ Phys.\ JETP {\bf 44}, 443 (1976), Sov.\ Phys.\ JETP {\bf 45}, 199 (1977); I.\thinspace I.~Balitsky and L.\thinspace N.~Lipatov, Sov.\ J.\ Nucl.\ Phys.\ {\bf 28}, 822 (1978), JETP\ Lett.\ {\bf 30}, 355 (1979).

\bibitem{Fadin:1998py} V.\thinspace S.~Fadin and L.\thinspace N.~Lipatov, Phys.\ Lett.\ B {\bf 429}, 127 (1998); G.~Camici and M.~Ciafaloni, Phys.\ Lett.\ B {\bf 430}, 349 (1998).

\bibitem{Andersen:2003an}
J.~R.~Andersen and A.~Sabio Vera, Phys. Lett. B, in press, hep-ph/0305236.

\bibitem{LLAlimit} J.~Kwiecinski, C.~A.~Lewis and A.~D.~Martin, Phys.\ Rev.\ D {\bf 54} (1996) 6664; C.~R.~Schmidt, Phys.\ Rev.\ Lett.\  {\bf 78} (1997) 4531; L.~H.~Orr and W.~J.~Stirling, Phys.\ Rev.\ D {\bf 56} (1997) 5875.

\bibitem{NLLpapers} D.A.~Ross, Phys. Lett. {\bf B431} (1998) 161;
Yu.V.~Kovchegov and A.H.~Mueller, Phys. Lett. {\bf B439} (1998)
423; J.~Bl\"umlein, V.~Ravindran, W.L.~van Neerven and A.~Vogt,
preprint DESY-98-036, {\tt hep-ph/9806368}; E.M.~Levin, preprint
TAUP 2501-98, {\tt hep-ph/9806228}; N.~Armesto, J.~Bartels,
M.A.~Braun, Phys. Lett. {\bf B442} (1998) 459; G.P.~Salam, JHEP
{\bf 8907} (1998) 19; M.~Ciafaloni and D.~Colferai, Phys. Lett.
{\bf B452} (1999) 372; M.~Ciafaloni, D.~Colferai and G.P.~Salam,
Phys. Rev. {\bf D60} (1999) 114036; R.S.~Thorne, Phys. Rev. {\bf
D60} (1999) 054031; S.~J.~Brodsky, V.~S.~Fadin, V.~T.~Kim, L.~N.~Lipatov and G.~B.~Pivovarov, JETP Lett.\  {\bf 70} (1999) 155; C.~R.~Schmidt, Phys.\ Rev.\ D {\bf 60} (1999) 074003; J.~R.~Forshaw, D.~A.~Ross and A.~Sabio Vera, Phys.\ Lett.\ B {\bf 455} (1999) 273; G.~Altarelli, R.~D.~Ball and S.~Forte, Nucl.\ Phys.\ B {\bf 575} (2000) 313.

\bibitem{running} L.N. Lipatov, {JETP}  ${\bf {63}}$, 904 (1986);
G. Camici and M. Ciafaloni,  {Phys. Lett.} B  ${\bf {395}}$, 118  (1997); R.~S.~Thorne, Phys.\ Lett.\ B {\bf 474} (2000) 372; J.~R.~Forshaw, D.~A.~Ross and A.~Sabio Vera, Phys.\ Lett.\ B {\bf 498} (2001) 149; R.~S.~Thorne, Phys.\ Rev.\ D {\bf 64} (2001) 074005; M.~Ciafaloni, M.~Taiuti and A.~H.~Mueller, Nucl.\ Phys.\ B {\bf 616} (2001) 349; M.~Ciafaloni, D.~Colferai, G.~P.~Salam and A.~M.~Stasto, Phys.\ Lett.\ B {\bf 541} (2002) 314, Phys.\ Rev.\ D {\bf 66} (2002) 054014.

\bibitem{phenomenology}
J.~R.~Forshaw and A.~Sabio Vera, Phys.\ Lett.\ B {\bf 440} (1998) 141; J.~R.~Forshaw, A.~Sabio Vera and B.~R.~Webber, J.\ Phys.\ G {\bf 25} (1999) 1511; J.~R.~Andersen, V.~Del Duca, S.~Frixione, C.~R.~Schmidt and W.~J.~Stirling, JHEP {\bf 0102} (2001) 007; J.~R.~Andersen, V.~Del Duca, F.~Maltoni and W.~J.~Stirling, JHEP {\bf 0105} (2001) 048; J.~R.~Andersen and W.~J.~Stirling, JHEP {\bf 0302} (2003) 018.

\end{thebibliography}
\end{document}